\documentstyle[12pt,epsfig]{article}
\newcommand{\be}{\begin{equation}}
\newcommand{\ee}{\end{equation}}
\newcommand{\bea}{\begin{eqnarray}}
\newcommand{\eea}{\end{eqnarray}}

\newcommand{\nn}{\nonumber}

\begin{document}
\begin{titlepage}

\title{Non-perturbative solution of metastable scalar models }

\author{Vladim\'{\i}r \v{S}auli\footnote{{ \it Email address:} sauli@ujf.cas.cz}\\
{\footnotesize \it Department of Theoretical Nuclear Physics, 
Nuclear Institute,'Re\v{z} near Prague}}

\maketitle

\begin{abstract}
Schwinger-Dyson equations for propagators 
are solved for the scalar $\Phi^3$ theory and massive   Wick-Cutkosky model.
 With the help of integral representation  the results
are obtained directly in Minkowski space in and beyond bare vertex approximation.
Various  renormalization scheme are employed which differ
 by the finite strength field renormalization function  $Z$ . 
The $S-$matrix is puzzled from the Green's function
and the  effect of truncation of the  DSEs is studied. Independently on the approximation
the numerical solution breaks down for certain critical value of the coupling constant, for
which the on-shell renormalized propagator starts to develop the unphysical singularity at very high space-like
square of momenta.\\ \\
{\it{PACS:}} {11.10.Gh, 11.15.Tk}\\  \\ 
{\it Keywords:} {Dyson-Schwinger equation, Renormalization, Spectral
representation}
\end{abstract}

\end{titlepage}

\section{Introduction}

The Dyson-Schwinger equations (DSE) are an infinite tower of coupled integral 
equation relating Green functions of the quantum field theory. If solved 
exactly, they would provide solutions of the underlying quantum field theory. 
In practice, the system of equations is truncated and one hopes to get some 
information, in particular on the solution in the non-perturbative regime,    
from solving the simplest equation for the two-point Green functions- the 
propagators. The other vertex functions, which also enter the DSE for the propagator,
are either taken in their bare form or some physically motivated Ansatze is employed.

In most papers dealing with the solution of DSEs, the Wick rotation from the Minkowski 
to Euclidean space is used to avoid singularities of the kernel inherent to the physical 
Greens functions. To our knowledge, the only exception is the series of papers 
\cite{SALAM},\cite{DELB},\cite{DELB1},\cite{DELB2} employing so-called "gauge technique" in quantum 
electrodynamic and its gauge invariant extension to quantum chromodynamic \cite{CORNWAL},
this work represents the born of the "Pinch Technique". Until now, the above-mentioned 
approach  has been never used in its non-perturbative context. Although not dealing with gauge 
theories, similarly to these techniques we instead solve the  directly in the momentum space,
 making use of the known analytical structure of the propagator, expressed
via the spectral decomposition. In the spectral or dispersive technique, we write  the Green function
as spectral integral over certain weight function and denominator parameterizing known
or assumed analytical structure. The generic spectral decomposition of the renormalized
 propagator reads:

\be
 G(p^2) = \int d \alpha \frac{\tilde{\sigma}(\alpha)}{p^2- \alpha- i \epsilon} \  ,
\label{genspec}
\ee 
where $\tilde{\sigma}(\alpha)$ is called Lehmann weight or simply the spectral function. 
If the threshold is situated above the particle mass, as it is for the stable 
(and unconfined) particles then the spectral function typically looks like 

\be
\tilde{\sigma}(\alpha)=r\delta(m^2-\alpha)+\sigma(\alpha)
\ee
where the singular delta function corresponds with non-interacting fields
and  $\sigma $ is appeared due to the interaction. Finite parameters  $r$ then represents the propagator
residuum and is simply related to field renormalization.
 It is also supposed that  $\sigma$ is a 
positive regular function which is spread smoothly from the zero at the threshold. 
Note here that the positivity of Lehmann weight is not required for our solution, 
but the models studied in this paper naturally embodied this property 
;see for instance \cite{SCHWEBER}, or any standard textbook.

Putting the spectral decomposition of the propagators and the expression for the vertex 
function into the DSE allows one to derive the real integral equation for the      
weight function $\sigma(\alpha )$. This equation involves only one real principal value 
integration  and can be solved numerically by iterations. 
Our solutions are obtained both for space-like and time-like propagator momenta; obtaining
this in the Euclidean approach would require  tricky backward analytical continuation.
Since all momentum integration are performed analytically,
there is no numerical uncertainty following from the renormalization which is usually present
in Euclidean formalism \cite{ROBERT}. Here, the renormalization procedure is performed
analytically with the help of the direct subtraction in momentum space. 
This perturbative perfectly known 
renormalization scheme (see \cite{COLLINS} for scalar models  and 
 \cite{SIVERS} for QED case, where a the comparison with other perturbative 
renormalization schemes was also made) has been already
applied to the QED and Yukawa model \cite{SAULI} in its non-perturbative context. 
In this paper, the off-shell
momentum subtraction renormalization scheme was introduced and used. 
In order to simplify the technique and 
 to compare various schemes we restrict ourselves to the choice of on mass-shell
 subtraction point $\mu=m$.

In this work, we would like to present  certain solutions of rather obscure 
theories: $\Phi^3$ and $\Phi_i^2\Phi_j$ scalar models.
 The second model will be referred here as the (generalized) Wick-Cutkosky model (WCM). 
In fact, not only models mentioned above but the all super-renormalizable four-dimensional scalar models  are not properly
defined since they have no true vacuum \cite{BAYM}. Instead of this they have   only metastable vacua
 (here we assume  non-zero masses of all particle content, in the opposite case the appropriate
 classical potential would not posses any local minimum).
 Instead of discarding these types of models, as sometimes happens, we  look
 whether this 'inconsistency' can be captured  by the formalism of  DSEs, or whether
 the appropriate solutions 'behaves ordinarily'.
The property of  super-renormalizability  makes our models particularly suitable for this purpose.
Actually, the super-renormalizability  here implies  the finiteness of the renormalization field constant $Z$ which 
therefore  can not be considered at all  (i.e.$Z=1$).
In the case of $\Phi^3$ theory we do not fully omit the field renormalization  
,but with the help of the  appropriate choice of the constant $Z$,
 we choose  the given  renormalization scheme.
Making this explicitly and after the evaluation of the scattering amplitude we look (in each scheme) at
whether 
the observables do converge (in all schemes) to some experimentally 
measurable values of the virtual scalar world .
As the suitable observables we choose
the amplitude $M$ for the scattering process  $\Phi\Phi\rightarrow \Phi\Phi$ 
and we have no find any  unexpected or even pathology behavior.
Instead of this,  when the approximation of the full solutions improve, 
we will see that the amplitudes calculated in the various renormalization schemes 
tends to converge to each other,  i.e. in this aspect, 
the $\Phi^3$ theory behaves as  the ordinary and physically meaningful one.
Here, this is the right place to note that the models with the metastable ground state serve
as an useful methodological tool, the role in which they are often employed.
In fact, $\Phi^3$ theory  serves as a good ground for the study of the various phenomena 
 \cite{CORNWALL5}, \cite{CORNWALL6}, \cite{DELBOUN},\cite{KREIMER}
(including phenomena like non-perturbative asymptotic freedom  and non-perturbative renormalization). 
There also exist a number of papers dealing with WCM. The DSEs 
for propagators of the WCM in their simple bare vertex approximation have been solved for the purpose of
calculation relativistic bound states \cite{AHLIG} (for other recent work dealing with
the bound states problem  within the WCM see
\cite{TJON} and references therein). For the purpose of  comparison with \cite{AHLIG}   we solve the
exactly analogical Minkowski problem. The obtained value of the critical coupling 
should  depends on the renormalization scheme. 
Having this slight dependence under  control it allows us to compare
with  other non-perturbative method \cite{ROSEN1},\cite{ROSEN2}. 
The comparison with conventional perturbation theory is  also made.

Regardless of the facts mentioned above, we are far from  concluding that $\Phi^3$   model
is a fully physically satisfactory one, since we do not know anything about the full solution.
 At this point, the study presented in this paper and the  studies
of the $\Phi^3$ model in five \cite{CORNWALL5} and six \cite{CORNWALL6} 
dimensions are conclusive in a similarly cautious way. Probably, a  more sophisticated conclusion
could be obtained by some lattice study, which has not yet done for this purpose.

At the end of the introduction, we should  mention  that there is always the possibility of including a 
sophisticated cut-off function $f(\Lambda)$ 
into the Lagrangian and regard our cubic models as an effective model bellow this cut-off. 
The theory at energies above $\Lambda$ could be another field theory or string theory, or whatever. 
However, this method is  developed and the appropriate
Polchinski renormgroup equations may be written down \cite{POLCHINSKI}, \cite{MORRIS}
these cut-off methods  lie somehow beyond the scope of this 
paper and we prefer to use the usual renormalization schemes,  where the independence on the appropriate 
regularization procedure is manifest.
Clearly, with the use of cut-off method it would be difficult to perform  the aforementioned comparison
 with the results \cite{AHLIG}, where on mass shell renormalization scheme has been performed.    
 Furthermore, we should note here that the dispersion technique used   
thorough the proposed paper would become more complicated due to the presence of the profile function   
$f(\Lambda)$.

In the next section we present the DSEs for $\Phi^3$ for propagator and vertex functions.
Subsequently we discuss 
the renormalization procedure and rewrite the propagator equation into its spectral form. 
Also, the numerical results 
and limitations are discussed.
The WCM is dealt with in the section 3. It is solved numerically in its  pure bare vertex approximation.
 The details of calculations are relegated
to the appendices A and B.

\section{ $  \Phi^3  $ theory }

\subsection{Dyson-Schwinger equation  for  $  \Phi^3  $ theory} 

The Lagrangian density for this model  reads

\be \label{lagr}
{\cal L}=\frac{1}{2}\partial_{\mu}\phi_0(x)\partial^{\mu}\phi_0(x)-
\frac{1}{2}m_0^{2}\phi_0^{2}(x)-g_0\phi_0^{3}(x)
\ee

where index $0$ indicates the unrenormalized  quantities. With the help of the functional 
differentiation of the generating functional
(for this procedure, see for instance \cite{ITZYKSON})
with classical action determined by (\ref{lagr}) one gets the following DSE 
(after transforming into the momentum space)
for the inverse propagator

\bea  \label{UDSE}
G_0^{-1}(p^2)&=&p^2-m_0^2-\Pi_0(p^2)
\nn \\
\Pi_0(p^2)&=&i3g_0 \int\frac{d^4q}{(2\pi)^4}\Gamma_0(p-q,q)G_0(p-q)G_0(q)
\eea 
where $\Gamma_0$ is the full irreducible  three-point vertex function
which satisfies its own DSE (\ref{dseforrg}). The integral of $\Pi_0$ is divergent and requires
the mass renormalization. Making the on-mass-shell subtraction  we define renormalized selfenergy ${\Pi_{R1}}$
 
\be
\Pi_{R1}(p^2)=\Pi_0(p^2)-\Pi_0(m^{2})
\ee
where $m$ is the pole "physical" mass, given by the equation $G^{-1}(m^2)=0$.
Defining the mass counter-term

\be
m^2=m_0^2-\delta m^2, \quad \delta m^2=\Pi_0(m^2)
\ee
and introducing additional finite renormalization constant

\be\label{konvence}
\phi_0= \sqrt{Z}\phi ; \quad 
g_0=g\frac{Z_g}{Z^{\frac{3}{2}}}
\ee
we obtain the inverse of the full propagator in term of physical mass

\bea \label{DSE2}
G^{-1}(p^2)&=&Z(p^2-m^2)-\Pi_1(p^2)
\nn \\
\Pi_1(p^2)&=&Z\left(\Pi_0(p^2)-\Pi_0(m^2)\right)
\nn \\
Z\Pi_0(p^2)&=&i3g^2 \int\frac{d^4q}{(2\pi)^4}\Gamma(p-q,q)G(p-q)G(q)
\eea 
where $g$ is a renormalized coupling and the constant $Z_g$ corresponds with the 
renormalization of  the vertex
function, and $G$ represents the  renormalized propagator with respect to the field strength
renormalization, i.e.

\be
\Gamma=Z_g\Gamma_0, \quad \quad G_0(p^2)=ZG(p^2)
\ee

We closed the system of our DSEs already at the level of equation for proper vertex.
Instead of solving the full renormalized DSE for the vertex

\be   \label{dseforrg}
g\Gamma(p,l)=6g+6g i\int\frac{d^4q}{(2\pi)^4}\Gamma(p,q)G(q)G(l-q)M(q,l,p)
\ee
we approximate the vertex by the first two terms of the appropriate skeleton expansion

\be\label{vert}
g\Gamma(p,l)=6g +i(6g)^3\int\frac{d^4q}{(2\pi)^4}G(q)G(p-q)G(l-q).
\ee
i.e., we approximate the vertex inside the loop by its bare value and 
the scattering matrix $M$ in Eq. (\ref{dseforrg})   is taken in its dressed tree approximation,i.e.,  $M=G$.
In the following, we will call the {\it dressed vertex} (DV) or improved approximation for the solution 
of the  propagator when the equation (\ref{vert}) is used for obtaining the triplet scalar vertex 
$\Gamma$ and  in the same spirit we use the name {\it bare vertex} (BV) for such a solution where only
the bare vertex was used. The improvement of the approximation is achieved by the skeleton expansion of the 
proper Green function where the series of DSEs is thrown away. Here,
 this is done at the level of the triplet vertex.
In \cite{DETMOLD}, we can see how the problem is becoming more complicated when 
$M$ is nontrivially taken into account. 
    
The equation  for the  propagator is solved in BV and DV approximations
 at each renormalization scheme separately.
 We  define these in the following section.

\subsection{Choosing the scheme}

We  assume (or rather we  neglect it) that there interaction does not create  
the bound states contributing to the weight function
$\sigma$. We use  the name  {\it minimal momentum subtraction} renormalization scheme (MMS)  the one
 where the only mass subtraction is used and where the field leaves unrenormalized, i.e. $Z=1$.
Therefore, we can write the spectral decomposition for the 
propagator  and for the self-energy $\Pi$ in the following form:

\bea \label{vrku}
G(p^2)&=& \frac{r}{p^2-m^2}+\int_{4m^2}^{\infty} d \alpha \frac{\sigma(\alpha)}{p^2- \alpha- i \epsilon}
\nn \\
&=&\left\{p^2-m^2-\Pi_1(p^2)\right\}^{-1}
\nn \\
\Pi_1(p^2)&=& \int_{4m^2}^{\infty}d\alpha
\frac{\rho(\alpha)(p^2-m^2)}{(\alpha-m^2)(p^{2}-\alpha+i\epsilon)} \quad ,
\eea
where $\pi\rho(s) $ represents the self-energy absorptive part 
and the threshold value of momentum $P_t^2=4m^2$ is explicitly written. 
Obviously, in this MMS the propagator does not have the pole residue $r$ equal to unity

\bea \label{reziduum}
&&\lim_{p^2\rightarrow m^2}(p^2-m^2)G(p^2)=\left[1-\frac{d}{d p^2}\Pi_1(p^2)|_{p^2=m^2}\right]=r
\nn \\
&&\frac{d}{d p^2}\Pi_1(p^2)|_{p^2=m^2}=
\int_{4m^2}^{\infty} d\alpha\frac{-\rho(\alpha)}{(\alpha-m^{2})^2} \quad.
\eea
After a simple algebra and taking the imaginary part of equation (\ref{vrku}) we arrive 
at the relation between the spectral functions $\sigma$ and $\rho$

\be    \label{symb}
\sigma(\omega)=\frac{r\rho(\omega)}{(\omega - m^2)^2}+
\frac{1}{\omega - m^2}
 P.\int_{4m^2}^{\infty} d\alpha \frac{\sigma(\omega)\rho(\alpha)\frac{\omega-m^2}{\alpha-m^2}
+\sigma(\alpha)\rho(\omega)}{\omega-\alpha}
\ee
where  $P.$ denotes principal value integration.

This is the first from two necessarily coupled equations which we actually solve for a given theory.
We discuss it in some details since its form depends only on the adopted renormalization procedure, 
not on the actual form of the interaction, 
nor on the approximation employed for the vertex function $\Gamma$ in the DSE for the propagator. 
The second equation connecting $\sigma$ and $\rho$ does depend on the form of vertex. Its derivation
is more complicated and we deal with it in the Appendix A.  

In some cases, the form (\ref{symb}) is not the most convenient one; 
for instance, when we want to look the bound state spectrum influence 
causes just by the self-energy effect (\cite{SAULI2}) .
Note the presence of the constant $r$ in the first term on the right-hand side,
it has to be determined from the relations (\ref{reziduum}) after each iteration. To get rid of this, we 
define the usual {\it on-shell renormalization scheme with unit residuum} (OSR scheme) by  

\be
Z=1+\delta Z
; \quad \delta Z=\frac{d}{d p^2}\Pi_1(p^2)|_{p^2=m^2}
\ee
which gives the standard receipt how to calculate the OSR propagator

\bea  \label{prt}
G^{OSR}(p^2)&=&\left\{p^2-m^2-\Pi_2(p^2)\right\}^{-1}
\nn \\
\Pi_{2}(p^2)&=&\Pi(p^2)-\Pi(m^{2})-\frac{d}{d p^2}\Pi(p^2)|_{p^2=m^2}(p^2-m^2)
\nn \\
\Pi(p^2)&=&i3g_2^2 \int\frac{d^4q}{(2\pi)^4}\Gamma(p-q,q)G(p-q)G(q)
\eea
and subsequently implies the spectral decomposition for $G^{OSR}$ and $\Pi_{2}$

\bea \label{brku}
G_{OSR}(p^2)&=& \frac{1}{p^2-m^2}+\int_{4m^2}^{\infty} d \alpha \frac{\sigma_2(\alpha)}{p^2- \alpha-i \epsilon}
\nn \\
\Pi_2(p^2)&=& \int_{4m^2}^{\infty}d\alpha
\frac{\rho_2(\alpha)(p^2-m^2)^2}{(\alpha-m^2)^2(p^{2}-\alpha+i\epsilon)}.
\eea

The relation between $\sigma_2$ and $\rho_2$ is now derived in the same way as before and it reads

\be    \label{symb2}
\sigma_2(\omega)=\frac{\rho_2(\omega)}{(\omega-m^2)^2}+\frac{1}{\omega-m^2}
P.\int_{4m^2}^{\infty} d\alpha\frac{\sigma_2(\omega)\rho_2(\alpha)\left[\frac{\omega-m^2}{\alpha-m^2}\right]^2
+\sigma_2(\alpha)\rho_2(\omega)}{\omega-\alpha}.
\ee

Note, that the Eqs. (\ref{symb}),(\ref{symb2}) are  inequivalent due to the scheme difference, the appropriate dependence   
of the weights $\rho$ and $\rho_2$  on the coupling constant $g $ and $g_2$ is explicitly written 
in  appendices A and  B, respectively.
( Two   inequivalent renormalization schemes should give the different Green function, but should give
the same S-matrix). 
 
At the end of this section, we very briefly discuss dimensional renormalization prescription 
\cite{DIMREG}, showing here that it is fully equivalent to MMS to all orders 
(note that the perturbation theory 
is naturally generated by the coupling constant expansion of the DSEs solution). For this purpose we choose
 the {\it modified minimal subtraction} $\bar{MS}$ scheme, noting that any 
other sort of schemes based on the dimensional regularization method would be treated in the same way.
Since the only  infinite contribution are affected when this renormalization is applied, therefore
the contribution with the dressed vertex (master diagram and so that) satisfies the unsubtracted dispersion relation
while for instance the one loop skeleton self-energy  diagram 
(in a fact the only one irreducible contribution which is infinite in four dimension) looks
(for space-like momenta) like

\be
\Pi_{\bar{MS}}^{[1]}(p^2)=\frac{18g^2}{(4\pi)^2}\int_{0}^{1} dx ln\left\{\frac{m^2-p^2x(1-x)}{\mu_{t'Hooft}}\right\}+\Pi(p^2)_{finite}
\ee
where $\Pi_{finite}$ represents the omitted  finite  terms which are not affected by 
dimensional renormalization at all (since they are finite to the all orders). 

The inverse of the full propagator reads in this scheme

\be
G_{\bar{MS}}^{-1}(p^2)=p^2-m^2(\mu_{t'Hooft})-\Pi_{\bar{MS}}^{[1]}(p^2)-\Pi(p^2)_{finite}.
\ee
Identifying the pole mass by equality  $G_{\bar{MS}}^{-1}(p^2=m^2_p)=0$
we simply arrive to the result

\be
G_{\bar{MS}}^{-1}(p^2)=p^2-m_p^2-\Pi_1(p^2)
\ee
where 

\bea \label{iden}
m_p^2=m^2(\mu_{t'Hooft})+\Pi_{\bar{MS}}^{[1]}(m_p^2)+\Pi(m_p^2)_{finite}
\nn \\
\Pi_1(p^2)=\Pi_{\bar{MS}}^{[1]}(p^2)-\Pi(p^2)_{finite}
-\Pi_{\bar{MS}}^{[1]}(m_p^2)-\Pi(m_p^2)_{finite}
\eea

Since the pole mass is renormgroup invariant quantity,
we see that $\bar{MS}$ scheme exactly corresponds with the one subtraction 
renormalization scheme,i.e. the  MMS.
Note here, that in renormalizable models such identification is  not so straightforward but always possible
\cite{SIVERS}. Of course, the appropriate identification is then  rather complicated.
 To conclude this section, we can see that the popular renormalization prescription like $MS$ or $\bar{MS}$ schemes
can be ordinarily used  in the non-perturbative context. 
At this point we disagree with the opposite statement of 
the paper \cite{TONIK}.

\subsection{Test of scheme (in-)dependence}

The physical observables should be invariant not only with respect to the choice 
of renormalization scale, but also with  respect to the choice of renormalization scheme. 
The first invariance is more then manifest in our approach, 
since all the quantities  used here are  the renormgroup invariants.
The second mentioned invariance is less obvious and, in fact, it is clear only for some very simple cases.
(The most simple case is  the tree-level  amplitude evaluation,
where the residua of the propagators may be exactly absorbed into the redefinitions 
of the coupling constants; but  
of course, in this case the renormalization is not required ).
In any reasonable renormalizable quantum field theory it is strongly
 believed that the obtained exact  Green functions must build the same $S$-matrix. 
In perturbation theory, we usually have several first terms of perturbation expansion 
and we hope that  they offer satisfactory description of the nature  when the 
"right" choice of renormalization scheme is made \cite{SIVERS},\cite{QCD}.
Furthermore, we should be aware that  the  possible sum of infinite many
terms of perturbation series should be regarded as an asymptotic one.
In fact, the application of  some sophisticated resummation technique is necessary 
in that case \cite{ITZYKSON},\cite{FISHER}.    
   
In DSE treatment we can talk about the level of DSEs system truncation  
instead of a  given coupling order. 
In the text bellow we describe a simple  possible procedure  how to see the 
improvements of physical observable
when  it is calculated within the improved truncation of DSEs.
For this purpose the  BV and  the DV solutions of DSEs in the both 
MMS and OSR schema are used to compose the  
same physically measurable quantity.

For our explanation we have   explicitly  choosen
 the matrix element $M$ of the elastic scattering process 
$\phi\phi \rightarrow \phi\phi$ which can be written 

\be \label{emko}
M(s,t,u)=\sum_{a=s,t,u}\Gamma G(a)\Gamma+...=
\sum_{a=s,t,u}(6g)^2G(a)+... \quad.
\ee
 The dots denote the neglected boxes and crossed boxes
contributions and the letters $s,t,u $ in (\ref{emko}) represent the usual Mandelstam variables
that satisfies $s+t+u=4m^2$, since now, the external particles are on-shell.

Using the notations introduced in the previous section,
 then the matrix $M$ in MMS scheme is calculated as

\be \label{aprox1}
M^{MMS}=\sum_{a=s,t,u}(6g)^2G(a), 
\ee
where the propagator is calculated through equations (\ref{vrku}),(\ref{reziduum}),(\ref{symb}).
 For OSR, the scattering matrix is composed like

\be \label{aprox2}
M^{OSR}=\sum_{a=s,t,u}(6g_2)^2G_{OSR}(a) 
\ee
where the propagator is calculated through equations
(\ref{prt}),(\ref{brku}) and (\ref{symb2}) and the  relations for $\rho$'s
are reviewed in the appendices A and B.

In the ideal case we would obtain 

\be \label{invariance}
M_{MMS}=M_{OSR}\quad,
\ee
which should be consequence of exact scheme independence. In reality, the Rel.(\ref{invariance})
 is not exactly fulfilled  due to the truncation of DSEs system. 
In what follows, we describe how to check the 
consistent condition (\ref{invariance})
and how to see the appropriate deviation numerically. 

Clearly, the equality should be valid in each kinematic channel separately. 
For instance,  choosing the   $t$-channel
for this purpose and comparing 
the pole part of matrices $M$ we obtain the relation between the coupling at each scheme

\be  
g_2^2=rg^2\quad, 
\ee
where $r$ is the residuum calculated from equation  (\ref{reziduum}).
This implies for us  that if we calculate the Green's functions in the 
OSR scheme  to compose the same S-matrix
the Green's functions in the  MMS scheme must be calculated with the coupling $g_2=\sqrt{r}g$.
Having the results for $\sigma$ and $\sigma_2$ extracted from the DSEs solved in the appropriate schemes,
we can compare imaginary parts of scattering matrices  $ M_{MMS}$ and $M_{OSR}$. 
Our approximation  (\ref{aprox1}),(\ref{aprox2}) implies 

\be 
g_2^2\sigma_2(\omega)=g^2\sigma(\omega) \label{pisek}.
\ee 
How accurately this equality is fulfilled at non-trivial regime $t>4m^2$ can be simply checked.
For this purpose we evaluate the integral (weighted) deviation $E_N$

\bea
E_N&=&\frac{\int\left[M_{OSR}(t)-M_{MMS}(t)\right]\frac{dt}{t^N}}
{ \int\left[M_{OSR}(t)+M_{MMS}(t)\right]\frac{dt}{t^N}}
\nn \\
&=&\frac{\int\left[\sigma_2(t)-\sigma(t)/r\right]\frac{dt}{t^N}}
{ \int\left[\sigma_2(t)+\sigma(t)/r\right]\frac{dt}{t^N}}\quad,
\eea
where the parameter $N$ serves us for adjusting the regime of momenta we are interested in.
A larger value of $N$ enhanced the threshold values of momenta while the ultraviolet modes 
are suppressed in that case.
We choose $N=0,1$ for the purpose of this paper.
 
Let us stress at the end of this section, that the next leading order of $M$ is  scheme invariant
and all the difference therefore follow from the remnant of the full DSEs solution. 
Hence only negligible deviation   is expected for small couplings. 
Also in general, the deviation $E_N$ should decrease when  considered
 approximations become more and more close to the full nonperturbative 
solution and it should principally vanishe for
the exact solution. In  other words, $E_N$ must decrease when  approximation (truncation of DSEs) improves.
The results obtained by the  above sketched method are reviewed in the next section.

\subsection{Results}

The integral equation for Lehmann weights have been solved numerically by the 
method of iteration. The appropriate solutions, obtained for several hundreds of 
mesh points and with the use of some  sophisticated integrator, have an accuracy of 
approximately one part of $10^4$ for reasonable value  ($\lambda<<\lambda_{crit}$) of the coupling strength 
$\lambda $ and   increase (up to several \%) when  $\lambda\simeq\lambda_{crit}$. 
The coupling strength  is defined as  dimensionless quantity

\be \label{lambda}
\lambda=\frac{18g^2}{16\pi^2 m^2}.
\ee
The critical value of  $\lambda$ is simply defined by the collapse of ( numerically sophisticated)
solution of the imaginary part DSEs. Before making a 
comparison of physical quantities we present the numerical results for the
Green's functions. In Fig.1 the so called dynamical mass 

\be
M(p^2)=G^{-1}(p^2)-p^2
\ee
of $\Phi^3$ theory boson is presented for various coupling strengths
in both renormalization schemes. The infrared details are displayed in Fig.2.
The dynamical mass is not directly  physically  observable since it is scheme
dependent from the definition, the exception is the pole mass
which is scheme independent and renormgroup invariant as well. It is interesting
that there are time-like values of square of momenta where the 
propagators behave almost  like free ones no matter how the coupling constant 
is strong. This happens somewhere around the point $p_f^2=6m^2$ for OSR scheme
and approximately at $p_f^2=20m^2$ for MMS scheme, which implies the physical irrelevance
of such a behavior (Of course, 
there are always differences within the absorptive parts
$\pi\rho$ which are ordinarily coupling constant dependent at these points). 
The appropriate relevance of propagator dressing is best seen when the dressed propagator
is compared with the free one $G=(p^2-m^2)^{-1}$. From the Fig.3 and Fig.4. 
we can see that  the propagator function is the  most sensitive with 
respect to the self-energy correction for  threshold momenta where 
these correction are enhanced about  one magnitude, while
they are largely suppressed for the above   mentioned  values of  momenta $p_f^2$. 
Note that nothing from these things can  be read from the purely Euclidean approach. 
The results presented up to now have been calculated in the bare vertex approximation,
the solution with vertex correction included will be discussed bellow. 
The appropriate bar vertex approximation  critical coupling value is  $\lambda_{crit}^{OSR}\simeq 3.5$ for OSR scheme 
and   $\lambda_{crit}^{MMS}\simeq 5$ for MMS scheme.
Their different values are not a discrepancy but the necessary consequence of 
the renormalization scheme dependence. 

Furthermore, in order to see the effect of self-consistency of DSE treatment 
we compare the DSE result with the perturbation theory in OSR scheme. 
From the Fig.5 we can see that the perturbation theory is perfectly 
suited method when  applied somewhere bellow the critical value of the coupling.
Therefore, the main goal of 
our solution is the information about  the domain of validity of given model.    

The issue of   vertex improvement by the one loop skeleton
diagram and its appropriate effect on the DSEs 
solution  and scattering matrix is discussed in the text bellow.
First let us note that the critical values of the couplings  decrease and we have
$\lambda_{crit}^{dressed vertex}\simeq \lambda_{crit}^{bare vertex}/2$ which is roughly
valid for  both the renormalization scheme employed. We return to the question of meaning
 $\lambda_{crit}$ when we will discuss the WCM.

 To make our comparison of proposed methods more  meaningful, we do not compare the Green's functions 
but rather wee look on the scattering amplitudes $M$ calculated in  both 
renormalization schemes obtained in  both truncations of DSEs.   
In  Fig.6 we compare the imaginary parts of scattering amplitudes 
$M$ at a given kinematic channel. The comparison is
made  in the  way  proposed and described in the previous section.
Henceforth, what are actually compared in this Figure are the Lehmann weights
$\sigma$'s of the MMS scheme calculated for certain $\lambda_{MMS}$ 
and the rescaled Lehmann weights $r\sigma_2$ calculated for the OSR scheme
with  the appropriate  coupling strength $\lambda_{OSR}=r\lambda_{MMS}$.
It is apparent that the lines for $\Im m M_{OSR}(r\lambda_{MMS},t)$ 
and $\Im m M_{MMS}(\lambda_{MMS},t)$
for solutions with dressed vertices 
are much close each other then the solution with bare vertices. This statement  is valid
 for all $t$ for a given theory characterized by its coupling constant (with $\lambda_{MMS}$ fixed).
This is true for all couplings $\lambda's$, the only--but 
not so striking-- exception is certain infrared excess for the 
  value of  couplings closed to the critical one. 
Of course, the worse numerical accuracy  play the role in strong coupling. 
Nevertheless, we can see that  when the approximation improves 
then  there is apparent signal for achieving the  renormalization scheme independence for all the
values of the coupling constant. 

In order to see aforementioned  quantitative improvement 
we have calculated the appropriate deviations $E_N, N=0,1$. 
The results for some larger value of the couplings
are presented in the Tab.1. The  corresponding difference becomes negligible when 
$\lambda$ decrease and approaches its  'perturbative' value.
For better orientation the  infrared details for three choices 
of the  coupling constants are also displayed in Fig.7.

\section{DSEs for the WCM}

The massive WCM is given by the following Lagrangian

\be \label{vcml}
 {\cal L} = \sum_i\frac{1}{2} \partial_{\mu} \Phi_{i}
 \partial^{\mu} \Phi_{i} - \sum_i \frac{1}{2} m_{i}^{2} \Phi_{i}^{2}
 +\left(\frac{g_{13}}{\sqrt{2}}  \Phi_{1}^{2} +\frac{g_{23}}{\sqrt{2}}
 \Phi_{2}^{2}\right) \Phi_{3} +C.P.
\ee
where $C.P.$ means the appropriate counter-term part.
Here we choose the second renormalization scheme employed in the previous section, i.e.  
the propagators of all three particles have the unit residua.
All the  definitions of counter-terms $\delta Z_i ,\delta m_i, \delta g_i$  
correspond with the OSR defined previously
but now for each particle separately. 
Furthermore, we adjust the couplings to be
\be
 g_{i3}=\frac{Z _{g_{i3}}}{Z_i Z_3^{\frac{1}{2}}} g_{{i3}_0},  \quad \quad i=1,2
\ee
such that

\be
 g_{13}=g_{23} .
\ee

The equal mass case $m_1=m_2$ was already solved \cite{SAULI2}
for purpose of studying the self-energy effect on the bound state spectrum.
Here we solved the unequal mass case

\be
 \frac{m_1}{m_2}=4 ;\quad  m_3=m_2
\ee
and compare the result with the Euclidean version of solution \cite{AHLIG}.
We restrict ourselves to the 
bare vertex approximation which is sufficient for comparison with \cite{AHLIG}.
Since all the derivation is rather straightforward we simply review the results.
The renormalized DSEs in bare vertex approximation read

\bea  \label{WCEDSE}
  G_{Ri}^{-1}\left( p\right) &=& p^2-m_i^{2}-\Pi_{i(2)}(p^2) \quad \quad i=1,2
\nn  \\
  G_{R3}^{-1}\left( p\right) &=& p^2-m_3^{2}-\Pi_{3(2)}(p^2)
\nn \\  
\Pi_i(p^2)&=& i2g^{2}\int \frac{d^{4}q}{\left( 2\pi \right) ^{4}}
\,G_{3 }\left( p-q\right) G_{i}\left( q\right) \quad \quad  \quad i=1,2
\nn \\
\Pi_{3}(p^2)&=& ig^{2}\int \frac{d^{4}q}{\left( 2\pi \right) ^{4}}\,
  \sum_{i=1,2}
  G_{i}\left( p-q\right) G_{i}\left( q\right)
\eea
where the bracketed index denotes the renormalization scheme employed,and the second index 
labels the particle associated with the appropriate field in the Lagrangian (\ref{vcml}).
All the propagators satisfy the Lehmann representation with unit residuum and all the 
proper function obeys the double subtracted dispersion relation  (\ref{brku}).
Henceforth, the appropriate spectral weights are related through the relations
 
\be    \label{symb3}
\sigma_{i}(\omega)=\frac{\rho_{i}(\omega)}{(\omega-m_i^2)^2}
 +\frac{1}{(\omega - m_i^2)}P.\int_{(m_j+m_k)^2}^{\infty} d\alpha
\frac{\sigma_i(\omega)\rho_{i}(\alpha)\left[\frac{\omega-m_i^2}{\alpha-m_i^2}\right]^2+
\sigma_i(\alpha)\rho_{i}(\omega)}{\omega-\alpha}  ,
\ee
where for the indices $i=1,2$ we have $j=2,3; k=3,3$ and 
for particle 3 the index $j=k=2$  since it label the lighter particle with the mass $m_2$.
The expression for the absorptive parts read

\bea  
\rho_{\pi_i}(\omega)&=&\frac{2g^2}{(4\pi)^2}
\left[B(m_i^2,m_3^2;\omega)
+\int_{0}^{\infty} d\alpha \left(B(\alpha,m_i^2;\omega)\sigma_3(\alpha)
+B(\alpha,m_3^2;\omega)\sigma_i(\alpha)\right)
\right.
\nn \\
&+&\left.\int_{0}^{\infty} d\alpha d\beta
B(\alpha,\beta;\omega)\sigma_3(\alpha)
\sigma_i(\beta)\right]
,\quad \quad i=1,2
\nn \\
\rho_{\pi_3}(\omega)&=&\sum_{i=1,2}\frac{g^2}{(4\pi)^2}
\left[\sqrt{1-\frac{4m_i^2}{\omega}}
+2\int_{0}^{\infty} d\alpha B(\alpha,m_i^2;\omega)\sigma_i(\alpha)\right.
\nn \\
&+&\int_{0}^{\infty}\left. d\alpha d\beta
B(\alpha,\beta;\omega)\sigma_i(\alpha)\sigma_i(\beta)\right].
\eea
where we freely integrate over the whole range of positive real axis leaving the information about 
the appropriate thresholds and subthresholds absorbed in the definition of the function $B$.

The above set of equations has been actually solved numerically. 
The main result for us is the appearance of the critical
coupling strength $\lambda_c\equiv g_c^2/(4\pi m_2^2)=0.12$ 
which rather accurately corresponds with the point where the renormalization 
constant $Z_2$ turns out to be negative. The appropriate dependence of the  renormalization constants $
Z_i$  is  presented in  Fig.8 for all three particle. 
The obtained critical value is in reasonable agreement with the one obtained by 
the Euclidean solution of DSEs system \cite{AHLIG}, where $\lambda_c=0.086$, as well as with the 
critical value $\lambda_c=0.063$ which was found using
a variational approach \cite{ROSEN1},\cite{ROSEN2}.

Furthermore, the existence of the critical coupling of OSR scheme can be seen from the
analytical formula for the inverse of propagator   

\be
G_i^{-1}=p^2-m_i^2-
\int_{0}^{\infty} d\alpha
\frac{\rho_i(\alpha)(p^2-m_i^2)^2}{(\alpha-m_i^2)^2(p^{2}-\alpha+i\epsilon)}
\ee
which implies that for the strong enough coupling  the Landau pole should appear, which  must arise when
the factor $L$ 
\be \label{factor}
L=[1-\int_{0}^{\infty}\frac {\rho(\alpha)}{(\alpha-m^2)^2}]
\ee
is negative (when it is just zero then the Landau 
pole is situated in space-like infinite, and for the positive 
$L$ this singularity never appears due to finiteness of the appropriate integral in (\ref{factor})).
For negative $L$ the propagator cannot satisfy   Lehmann representation at all and at least 
the Minkowskian treatment used in this work must fail. 
Comparing equation (\ref{factor}) with the definition of renormalization constant $Z$
we clearly have the identification $L=Z$. 
As we have mentioned, the numerical solution start to 
fail when the condition $Z=0$ is fulfilled.  This statement is justified
with  10 \% numerical accuracy. 
(We have no similar guidance for MMS scheme but we expect the similar appearance of the critical coupling
$\lambda_{MMS}$ for this scheme as occured in $\Phi^3$ theory, 
but the reason for the numerical failure 
in this case is not found in this paper.)

\section{Conclusion}

We have obtained numerical solutions of the   DSEs in  Minkowski space for 
$ \Phi^3$ theory and the WCM. This suggests that the expansion of the theory
 around the metastable vacuum leads to the predicative result.
Our technique allows us
to extract propagator spectral function $\rho(s)$ with reasonably high numerical accuracy.
Since the renormalization procedure is performed analytically, it has no effect
on the precision of solution.  When the coupling does not exceed a certain critical value,
then the  domain of analyticity of the propagator is the all real axis of  $p^2$. 
An attempt to clarified the meaning of critical coupling value was made. This suggests 
that it  corresponds with  appearance of  unphysical singularity in the on-shell renormalized propagator.  
Consequently, the field renormalization constant (in on shell scheme) turns to be negative
for $\lambda>\lambda_{crit}$.

\appendix
\section{ Dispersion relations for self-energies in bare vertex approximation }

In this Appendix we derive DRs for self-energies in  both renormalization schemes
for the bare vertex. The calculation is very straightforward, and in fact it represents nothing 
else but evaluation of the one loop scalar 
Feynman diagram with different masses in internal lines. 
 
Substituting the Lehmann representation for MMS propagators (\ref{vrku})
the unrenormalized $\Pi$ can be split like   

\bea  \label{SELF}
\Pi(p^2)&=&\Pi_{(b,b)}(p^2)+2\Pi_{(b,s)}(p^2)+\Pi_{(s,s)}(p^2)
\nn  \\
\Pi_{(b,b)}(p^2)&=&\int d\bar{q}
\frac{18r^2g^{2} }{((p+q)^{2}-m^{2}+i\epsilon)(q^{2}-m^{2}+i\epsilon)}
\nn\\
\Pi_{(b,s)}(p^2)&=&\int d\bar{q}
\int d\alpha\frac{18rg^{2}\sigma(\alpha)}
{(q^{2}-\alpha+i\epsilon)((p+q)^{2}-m^{2}+i\epsilon)}
\nn \\
\Pi_{(s,s)}(p^2)&=&\int d\bar{q}
\int d\alpha d\beta
\frac{18g^{2}\sigma(\alpha)\sigma(\beta)}
{((p+q)^{2}-\alpha+i\epsilon)(q^{2}-\beta+i\epsilon)},
\eea
where we have  used shorthand notation for the measure  $id^4q/(2\pi)^4\equiv d\bar{q}$.
Making the subtraction, we immediately arrive for the pure perturbative contribution
(up to the presence of the square of residuum):
 
\bea   \label{PP}
\Pi_{1(b,b)}&=&\int_{4m^2}^{\infty} d\omega \frac {\rho_{1(b,b)}(p^2-m^2)}
{(p^2-\omega +i\epsilon)(\omega-m^2)}
\nn \\
\rho_{1(b,b)} (\omega)&=&\frac{18r^2g^2}{(4\pi)^2}
\sqrt{1-\frac{4m^{2}}{\omega}}
\eea

The most general integral to be solved is similar to  the above case but with the physical masses  replaced
by the  spectral variables. 
 
\be  \label{gener}
I(p^2)=\int d\bar{q}
\frac{1}{((p+q)^{2}-\alpha)(q^{2}-\beta)}
\ee
which after the subtraction (\ref{firstline}) and integration over the Feynman parameter $x$
leads to the appropriate single-subtracted DR (\ref{gener2})

\bea    \label{firstline}
I_{1s}(p^2)&=&I(p^2)-I(m^2)= \frac{1}{(4\pi)^2}\int_{0}^{[1]}dx
\int_{\frac{m^{2}x+\alpha(1-x)}{x(1-x)}}^{\infty}
 d\omega\frac{(p^2-m^2)}
{(\omega-m^2)(p^{2}-\omega+i\epsilon)}
\nn \\
&=&\int_{0}^{\infty}d\omega \frac{p^2-m^2}{(p^2-\omega+i\epsilon)}  
\frac{B(\alpha,\beta;\omega)}
{(4\pi )^{2}(\omega-m^{2})} \label{gener2},
\eea
where the function $B(u,v;\omega)$ is defined through the  Khallen triangle function $\lambda$ 
like

\bea
B(u,v,\omega)&=&\frac{\lambda^{1/2}(u,\omega,v)}{\omega}
\Theta\left(\omega-(\alpha^{\frac{1}{2}}+\beta^{\frac{1}{2}})^2\right)
\nn \\
\lambda(u,\omega,v)&=&(u-\omega-v)^2-4\omega v
\nn \\
&=&\omega^2+u^2+v^2-2\omega v -2\omega u -2u v.
\eea
It can be easy checked that $B(m^2,m^2;\omega)=\sqrt{1-\frac{4m^2}{\omega}}\Theta(\omega-4m^2)$
which was already introduced in (\ref{PP}).

The OSR scheme requires additional subtraction which is finite and henceforth  
can  proceed by making a simple algebra

\bea \label{gener3}
I_{2s}(p^2)&=&I_{1s}(p^2)-\frac{d}{d p^2}|_{p^2=m^2}I_{1s}(p^2)
\nn \\
&=&\frac{1}{(4\pi)^2}\int_{0}^{\infty}d\omega \frac{(p^2-m^2)^2}{(\omega-m^{2})^2(p^2-\omega+i\epsilon)}
B(\alpha,\beta;\omega).
\eea

To summarize the results we see that MMS self-energy satisfies one subtracted DR
with the absorptive part $\pi\rho_1$ given like 

\bea  \label{fresult}
\rho_{\pi_1}(\omega)&=&\rho_{1(b,b)}(\omega) +2\rho_{1(b,s)}(\omega)+\rho_{1(s,s)}(\omega)
\nn \\
\rho_{1(b,b)}(\omega)&=&\frac{18r^2g^2}{(4\pi)^2}
\sqrt{1-\frac{4m^{2}}{\omega}}
\nn \\
\rho_{1(b,s)}(\omega)&=&\int_{9m^2}^{\infty}d\alpha
\frac{18rg^{2}}{(4\pi )^{2}}B(\alpha,m^2;\omega)\sigma(\alpha)
\nn \\
\rho_{1(s,s)}(\omega)&=&\int_{16m^2}^{\infty}d\alpha d\beta
\frac{18g^2}{(4\pi )^{2}} B(\alpha,\beta;\omega)\sigma(\alpha)\sigma(\beta)\quad,
\eea
while the self-energy in OSR scheme satisfies double subtracted DR with
the absorptive part $\pi\rho_2$

\bea  \label{sresult}
\rho_{\pi_2}(\omega)&=&\rho_{2(b,b)}(\omega) +2\rho_{2(b,s)}(\omega)+\rho_{2(s,s)}(\omega)
 \nn \\
\rho_{2(b,b)}(\omega)&=&\frac{18g^2}{(4\pi)^2}
\sqrt{1-\frac{4m^{2}}{\omega}}
\nn \\
\rho_{2(b,s)}(\omega)&=&\int_{9m^2}^{\infty}d\alpha
\frac{18g^{2}}{(4\pi )^{2}}B(\alpha,m^2;\omega)\sigma_2(\alpha)
\nn \\
\rho_{2(s,s)}(\omega)&=&\int_{16m^2}^{\infty}d\alpha d\beta
\frac{18g^2}{(4\pi )^{2}} B(\alpha,\beta;\omega)\sigma_2(\alpha)\sigma_2(\beta).
\eea


\section{Two-loop skeleton self-energy DR}

The finite  two-loop integral appears after the substitution of the vertex (\ref{vert})
to the self-energy formula

\bea \label{twopi}
\int d\bar{q}d\bar{k}
&&\left[ (k^2-\alpha_1+i\epsilon)((p+k)^2-\alpha_2+i\epsilon)
((k-q)^2-\alpha_3+i\epsilon)\right.
\nn \\
&&\left.(q^2-\alpha_4+i\epsilon)
((p+q)^2-\alpha_5+i\epsilon)\right]^{-1}
\eea
where all the irrelevant pre-factors are omitted for purpose of the brevity.
 They will be correctly added at the end
of  calculation for both renormalization schemes separately.
The contribution is ultraviolet finite therefore we first calculate the unrenormalized result.
Firstly,  we parameterize the  off-shell vertex  by matching the first three denominators 
and consequently we integrate over the  triangle  loop momentum $k$

\bea \label{vertex}
&&\int \bar{k}
\left[ (k^2-\alpha_1+i\epsilon)((p+k)^2-\alpha_2+i\epsilon)
((k-q)^2-\alpha_3+i\epsilon)\right]^{-1}=
\nn\\
&&\int \bar{k}
\int_0^1 dx dy 2y \left[k^2xy+(p+k)^2(1-x)y+(k-q)^2(1-y)\right.
\nn \\
&&-\left.\alpha_1 xy-\alpha_2(1-x)y-\alpha_3(1-y)+i\epsilon\right]^{-3}=
 \\
&&\int_0^1 \frac{dx dy y}{(4\pi)^2}
\left[p^2(1-x)y(1-(1-x)y)+q^2y(1-y)+2p.q(1-x)y(1-y)\right.
\nn \\
&&\left.-\alpha_1 xy-\alpha_2(1-x)y-\alpha_3(1-y)+i\epsilon\right]^{-1}.
\nn
\eea
Next, we substitute $x\rightarrow 1-x$ and  
after some algebra we obtain for equation (\ref{vertex})

\be \label{form}
\int_0^1 \frac{dx dy }{(4\pi)^2(1-y)}
\left[q^2+2p.qx+p^2\frac{x(1-xy)}{(1-y)}-O_{1-3}+i\epsilon\right]^{-1}
\ee
where we have used short notation $O_{1-3}=\alpha_1 \frac{1-x}{1-y}
+\alpha_2\frac{x}{1-y}+\alpha_3\frac{1}{y}$.
We continue by matching equation (\ref{form}) with   
two spare denominators in (\ref{twopi})  by using Feynman  
variables $z $ and $u$ for denominators with $\alpha_4$ and $\alpha_5$, respectively.
Then we can write for (\ref{twopi}) 

\bea
&&\int d\bar{q} \int_0^1 \frac{dx dy dz du 2u }{4\pi)^2(1-y)}
\left[q^2+2p.qxzu +2p.q(1-u)\right.
\nn \\
&&+\left.p^2\frac{(1-xy)xzu}{(1-y)}+p^2(1-u)-O_{1-5}+i\epsilon\right]^{-3}
\eea
where we have used shorthand notation $O_{1-5}=O_{1-3}zu+\alpha_4(1-z)u+
\alpha_5(1-u)$. Shifting $q+p(xzu+1-u) \rightarrow q$ and integrating
over new $q$ it yields:

\bea
&&\int_0^1 \frac{dx dy dz du u }{(4\pi)^4(1-y)F(x,y,z)}
\frac{1}{\left[p^2-\frac{O_{1-5}}{F(x,y,z)}+i\epsilon\right]}
\nn \\
&&F(x,y,z)=1-u+\frac{(1-xy)xzu}{(1-y)}-(xzu+(1-u))^2
\eea 
Next we  substitute $u\rightarrow \omega$
where $\omega= O_{1-5}/F(x,y,z)$. Using the notation 

\bea \label{blabla}
\omega&=&\frac{u a_1+a_2}{u^2b_1+ub_2}
\\
a_1&=&(\alpha_1 (1-x)y
+\alpha_2 xy+\alpha_3 (1-y))z+(\alpha_4(1-z)-\alpha_5) y(1-y)
\nn \\
a_2&=&\alpha_5 y(1-y)
\nn \\
b_1&=&-(1-xz)^2y(1-y)
\nn \\
b_2&=&(1-2xz)y(1-y)+(1-xy)xyz\quad,
\eea
we can write down the appropriate DR for equation (\ref{twopi}) 

\bea  \label{rezek}
\Omega(\omega;\alpha_1,..\alpha_5)&=&\int_0^{\infty} \frac{d\omega}{p^2-\omega+i\epsilon}
\int_0^1\frac{dx dy dz}{(4\pi)^4(1-y)}\frac{\Theta\left(\omega-\frac{a_1+a_2}{b_1+b_2}\right)
\Theta(D)}
{\left[\frac{\alpha_5}{U^2}-\omega(1-xz)^2\right]}
\\
U&=&\frac{-B+\sqrt{D}}{2A}; \quad D=B^2-4AC
 \nn \\
A&=&\omega b_1; \quad B=\omega b_2-a_1; \quad C=-a_2 \quad.
\nn
\eea

Note here that spectral function  (everything after the first fraction in (\ref{rezek})
is always positive for allowed values of $\alpha's$
and it is regular function of its argument $\omega$. 
The various subthresholds are then given by the values of Lehmann variables $\alpha$'s
 in accordance with the step function presented, noting that the perturbative threshold is given
again by $4m^2$ and in that case case the result partially simplified.
  For completeness we reviewed the associated simplifications, namely:
 $a_1=m^2z(1-y(1-y)); \quad a_2=m^2y(1-y) $.
Making one subtraction for the MMS and two subtraction for OSR scheme
we can recognize that the appropriate skeleton DR for master diagram
has the absorptive part

\be  \label{uno}
\rho_{1}^{[2]}(\omega)=\frac{(6g)^4}{2}\prod_{i=1}^5 \int d\alpha_i 
\tilde{\sigma}(\alpha_i)\Omega(\omega,\alpha_1,..\alpha_2)
\ee
for MMS scheme and

\be
\label{duo}
\rho_{2}^{[2]}(\omega)=\frac{(6g_2)^4}{2}\prod_{i=1}^5 \int d\alpha_i 
\tilde{\sigma}_2(\alpha_i)\Omega(\omega,\alpha_1,..\alpha_2)
\ee
for OSR scheme, respectively.
In fact it  gives rise 28  various
contributions to $\rho^{[2]}$  (only 12 are actually topologically independent,
 distinguished by the number of continuous Lehmann weights with the appropriate position of spectral variable 
in $\Omega$.) All of them have been found numerically for the purpose of DSEs solution.

\newpage


\begin{center}
\small{\begin{tabular}{|c|c|c|c|c|}
\hline \hline
  &  $E_{N=0}$ BV  &$E_{N=0}$ DV  &$ E_{N=1}$ BV 
& $E_{N=1}$ DV  \\
\hline \hline
$\lambda=0.1  $& 0.020 & 0.0039  & 0.022 & 0.0078 \\
\hline
$\lambda=0.5 $ & 0.076 & 0.025  &  0.071 & 0.025   \\
\hline
$\lambda=1.0$ &  0.15 & 0.040 & 0.13      & 0.08   \\
\hline \hline
\end{tabular}}
\end{center}

TABLE 1. Normalized and weighted integral deviations $E_N$ between  scattering amplitudes calculated in the  OSR and MMS scheme.
Exact scheme independence corresponds to the case $E_N=0$.
The parameters $N=1$ makes the quantity $E$ more sensitive to the systematic error in the infrared domain.   
The denotation BV(DV)  means that 
the  appropriate propagator was calculated with bare (dressed) vertex. 
The function $E$ is displayed for three cases of results  presented in Figures 6 and 7.    


\newpage

\begin{figure}
\centerline{ \mbox{\psfig{figure=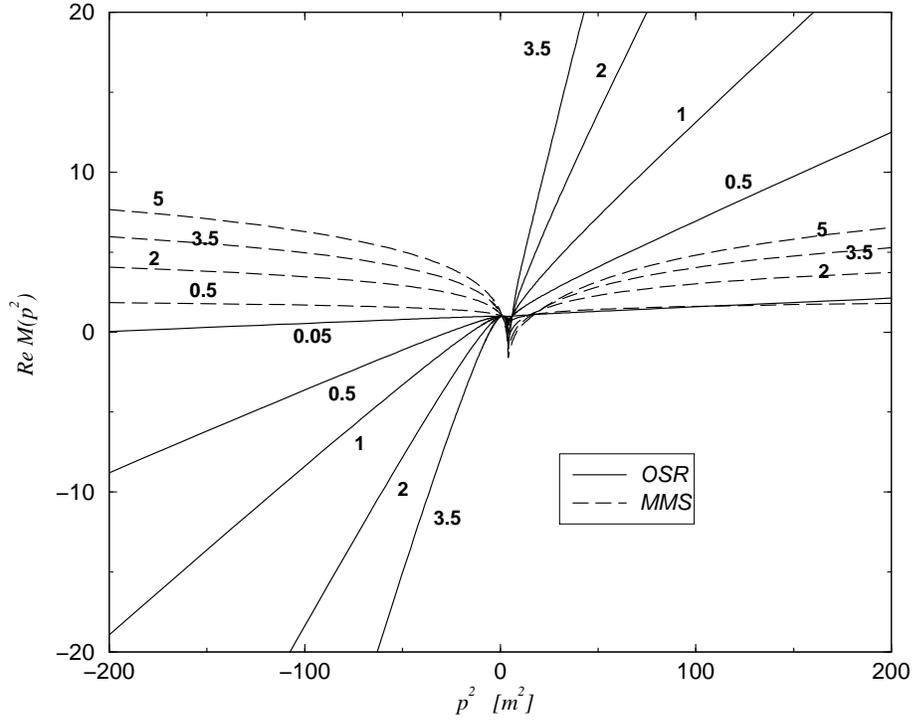,height=12truecm,angle=270}} }
\caption{Dynamical mass of scalar particle in $\Phi^3$ theory calculated 
in bare vertex approximation in the both renormalization schemes.
The lines are labeled by the value of
$\lambda_{MMS}$ for MMS scheme and $\lambda_{OSR}$ for OSR renormalization scheme.}
\end{figure} 

\begin{figure}
\centerline{ \mbox{\psfig{figure=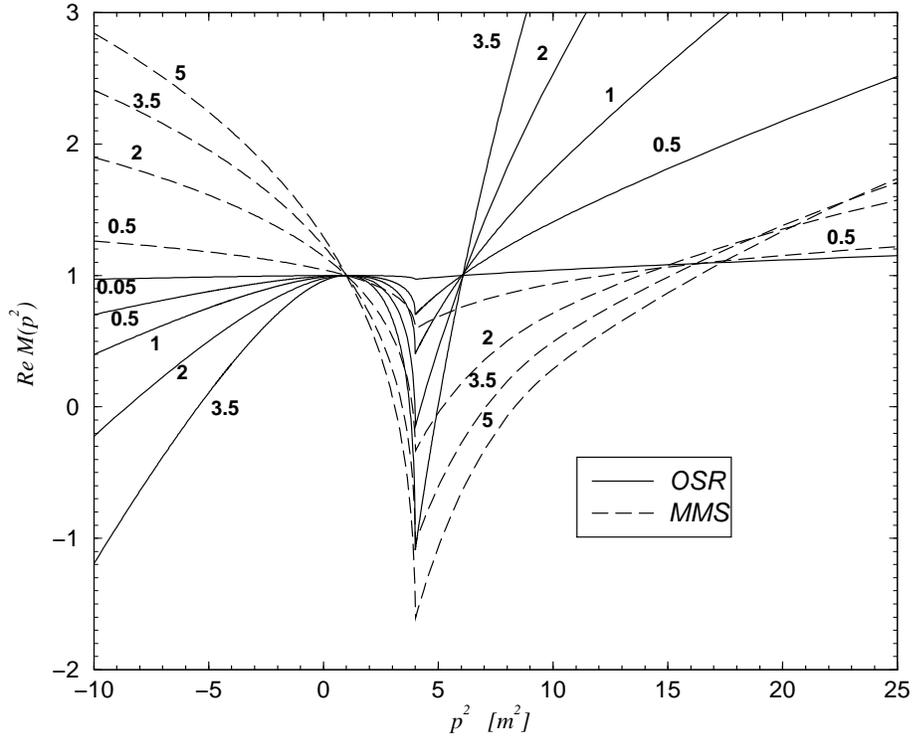,height=12truecm,angle=270}} }
\caption{ Infrared (threshold) details of the Fig.1.  }
\end{figure}

\begin{figure}
\centerline{ \mbox{\psfig{figure=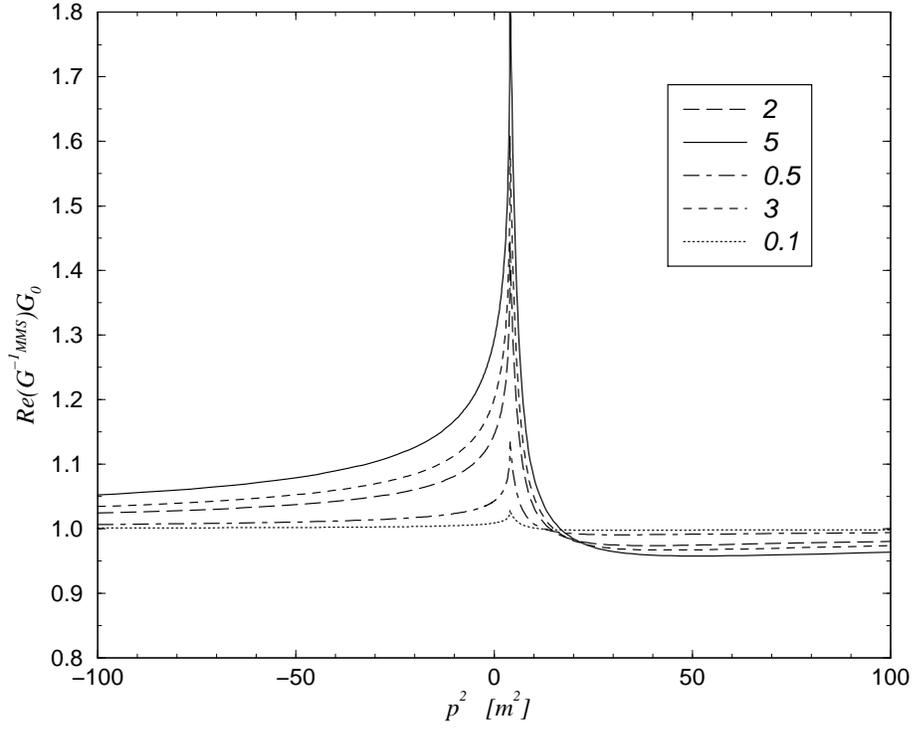,height=12truecm,angle=270}} }
\caption{ The propagators deviations from free theory. The propagator is calculated 
in minimal momentum  renormalization scheme   for various $\lambda_{MMS}$ }
\end{figure}

\begin{figure}
\centerline{ \mbox{\psfig{figure=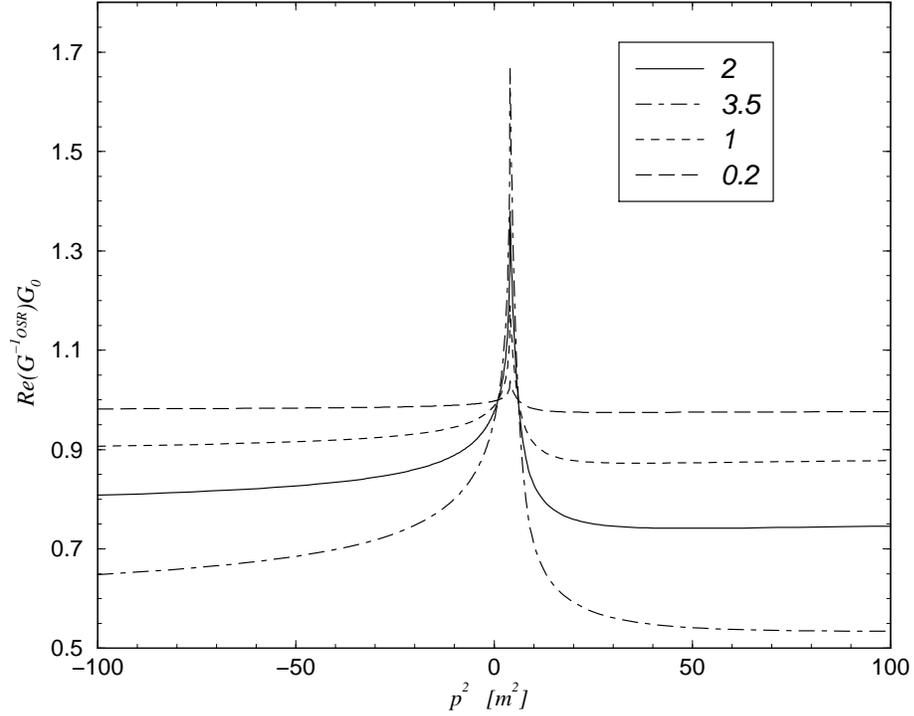,height=12truecm,angle=270}} }
\caption{ The propagator deviation from free theory. The propagator is calculated 
in on mass-shell renormalization scheme with unit residuum  for various $\lambda_{OSR}$. }
\end{figure}

\begin{figure}
\centerline{ \mbox{\psfig{figure=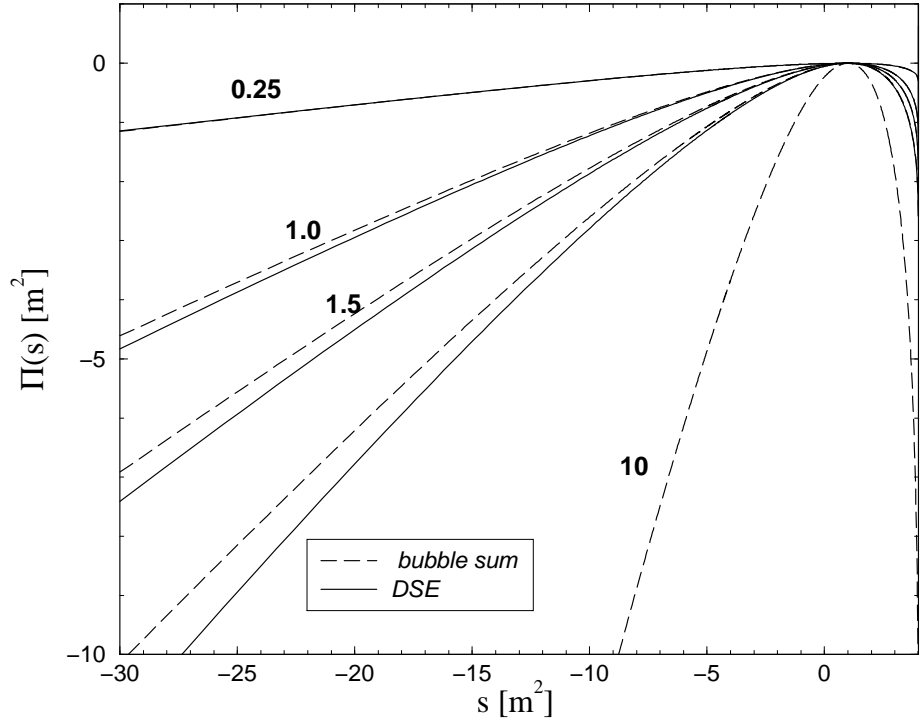,height=12truecm,angle=270}} }
\caption{ Comparison of DSE results in bare vertex approximation with the  perturbation theory result. DSE  and
bubble summation is compared in OSR scheme. Each two close lines off different types
 correspond to the same value of coupling $\lambda_{OSR}=\{0.25;1.0;1.5;2.2\} $  . 
The lowest dashed line with  $\lambda=10$  has not its DSE partner solution
(since $ \lambda>\lambda_c$).} 
\end{figure}

\begin{figure}
\centerline{ \mbox{\psfig{figure=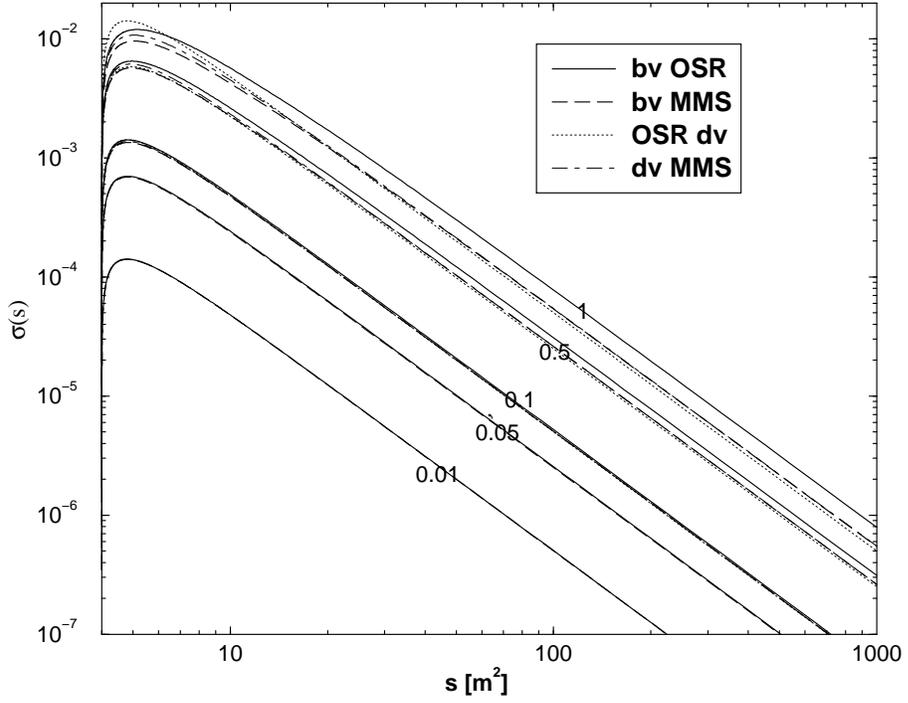,height=12truecm,angle=270}} }
\caption{ Imaginary parts of scattering matrix calculated with propagator
which have been obtained in MMS and OSR scheme with (dv) and without (bv) improved vertex.
Each set of lines corresponding to the same model is labeled by the  coupling strength $\lambda_{MMS}$.}
\end{figure}

\begin{figure}
\centerline{ \mbox{\psfig{figure=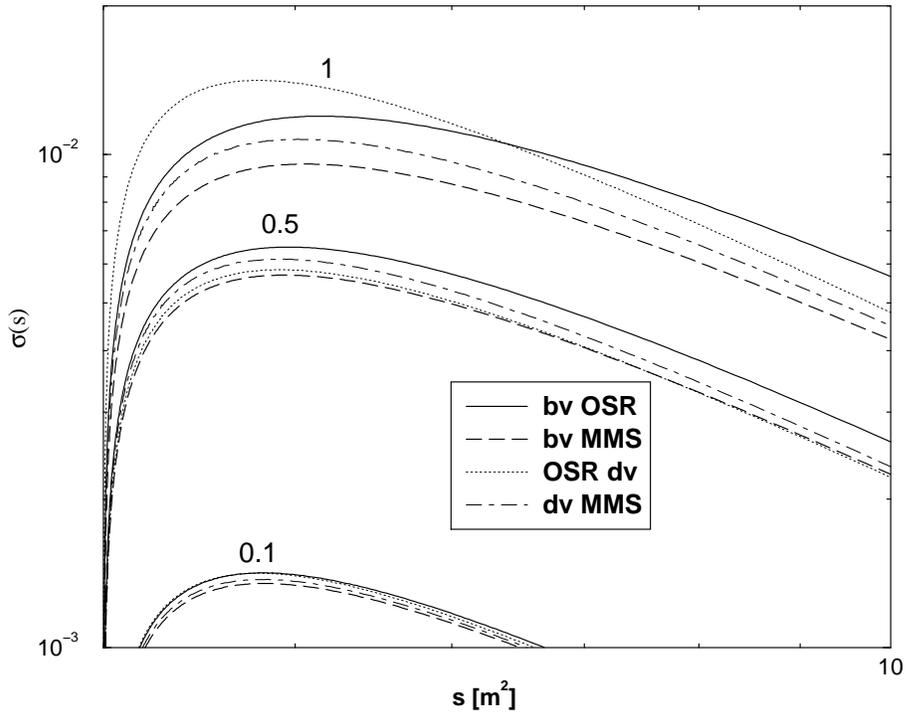,height=12truecm,angle=270}} }
\caption{ The low frequency details of the Fig.6.}
\end{figure}

\begin{figure}
\centerline{ \mbox{\psfig{figure=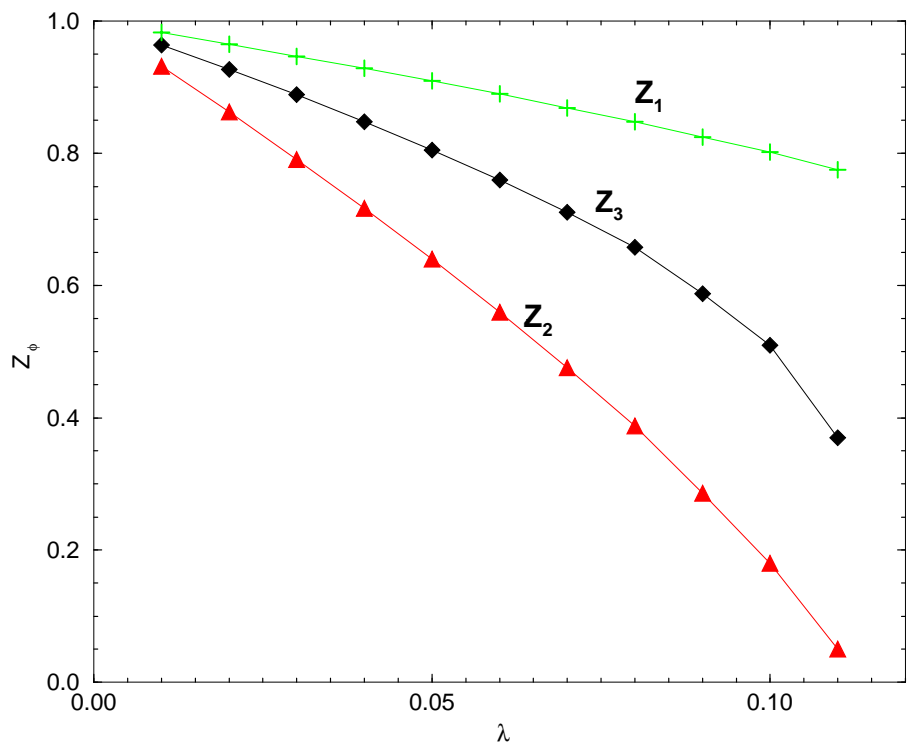,height=12truecm,angle=270}} }
\caption{ The dependence of field strength renormalization constants on the coupling strength of Wick-Cutkosky model.
The index 1-3 labels the particle. }
\end{figure}

\end{document}